\documentclass{article}
\usepackage{amsmath,amssymb,amsthm}
\usepackage{epsfig}
\newtheorem{theorem}{Theorem}  
\newtheorem{lemma}{Lemma} 
\newtheorem{corollary}{Corollary} 
\newcommand{\dd}{{\rm d}}

\begin{document}

\title{\bf \LARGE A finiteness proof for the\\ Lorentzian 
state sum spinfoam model\\ for quantum general relativity}

\author{\bf Louis Crane\,${}^{ab}$\footnote{e-mail: crane@math.ksu.edu}\ ,
Alejandro Perez\,${}^{cd}$\footnote{e-mail: perez@cpt.univ-mrs.fr}\ , 
Carlo Rovelli\,${}^{cd}$\footnote{e-mail: carlo@rovelli.org}\\[.1cm] 
\em ${}^a$ Mathematics Department, Kansas State University, USA \\ 
\em ${}^b$Istituto Superior Tecnico, Lisboa, Portugal\\
\em ${}^c$ CPT, Case 907, F-13288 Marseille, France\\
\em ${}^d$ Department of Physics, University of Pittsburgh, PA15260,
USA} 
\date{\today} 
\maketitle 
\begin{abstract} 
    We show that the normalized Lorentzian state sum is finite on any
    triangulation.  It thus provides a candidate for a perturbatively
    finite quantum theory of general relativity in four dimensions
    with Lorentzian signature.
\end{abstract}

\section{Introduction} 

In \cite{Barrett:1998gw}, a state sum, or spinfoam model, for
discretized Euclidean signature general relativity was proposed.  The
model was made finite by passing from the representations of $so(4)$,
the Euclidean Lorentz algebra, to its q-deformation at a root of
unity.  In \cite{Perez:2001fs}, a better motivated normalization for
the model in \cite{Barrett:1998gw} was derived, and it was conjectured
that with this new normalization the model was finite on any finitely
triangulated manifold without passing to the quantum group.  The
finiteness conjecture was proved in \cite{Perez:2001bf}.  The same
normalization was obtained, from a different perspective, in
\cite{Oriti:2001hh}. 

In \cite{Barrett:2000qw}, a Lorentzian signature version of the state
model in \cite{Barrett:1998gw} was proposed.  It was suggested that
the model could be made finite by passing to the representations of
the quantum Lorentz algebra \cite{Buffenoir:2000vd}.  In
\cite{Perez:2001ec}, a change of normalization, similar to the one
that made finite the Euclidean model, was proposed for the Lorentzian
model.  The purpose of the present paper is to prove that the
Lorentzian signature model of \cite{Barrett:2000qw} with the
normalization of \cite{Perez:2001ec} is in fact {\em finite\/} for any
finite triangulation of a 4-manifold.

This result is remarkable.  The model is an integral over infinite
domains of terms each of which is the trace of an operator on certain
infinite dimensional representations of the Lorentz algebra, the
balanced unitary ones.  If we adopt the point of view of
\cite{Perez:2001ec} that the sum on a specific triangulation should be
interpreted as a term in a perturbation series
\cite{DePietri:2000bx,Reisenberger:2000fy,Reisenberger:2001zc}, this
result implies that --up to the issue of singular triangulations
discussed below-- the theory is finite order by order, since there are
only a finite number of triangulations with a given number of
4-simplices.  For a theory of four dimensional Lorentzian quantum
general relativity, perturbative finiteness is significant. 
Finiteness on nondegenerate triangulations is proven here with
mathematical rigor.  The proof relies on technology developed in
\cite{Baez:2001fh}.

At the moment our proof does not work for degenerate triangulations,
and we do not know if these give finite contributions or not.  This
will require a more delicate analysis of the relevant integrals.  In
the Euclidean context, singular triangulations are finite
\cite{Perez:2001bf}.

The finiteness of the model is the result of several mathematical
surprises.  The finiteness of the individual 10J symbols comes about
by representing them as multiple integrals on hyperbolic space, then
making the extremely simple regularization of omitting any single
integration.  This regularization does not break any of the
symmetries.  This procedure does not work for spin nets for arbitrary
unitary representations of the Lorentz algebra, but only for the
balanced ones \cite{Barrett:2000qw}.  The finiteness of the complete
state sum on any finite triangulation, which we prove in this paper,
depends then on a rather delicate relationship between asymptotic
estimates for the relativistic spin nets which go into it.  

We recall the definition of the model in Section 2, we prove its
finiteness in Section 3, and discuss the result in Section 4.

\section{The model}

We do not review any of the derivations or the motivations of the
model here, but merely recall its definition.  We refer the reader to
the bibliography for a thorough introduction.

The principal series unitary representations of the Lorentz algebra
$so(3,1)$ are denoted $R(k, \rho)$, where $k$ is an integer and $\rho$
is a nonnegative real number.  The balanced representations are those
with $k=0$; they are just labelled with $\rho$.

We construct the state sum model by taking a (fixed) nondegenerate
finite triangulation of a 4-manifold, with or without boundary.  We
label each 2-simplex $f$ of the triangulation with a balanced
representation of the Lorentz algebra, or, more simply, with a
positive real parameter $\rho_{f}$.  The state sum is then given by
the expression
\begin{equation}
\label{one}
{\cal Z}= \int \limits^{\infty}_{\rho_{f}=0} \dd \rho_f \ \ 
\prod_{f} \rho^2_{f} \ 
\prod_{e} \Theta_{4}(\rho^e_{1}\dots \rho^e_{4}) \ 
\prod_{v} I_{10}(\rho^v_{1}\dots\rho^v_{10}). 
\end{equation}
The integration is over the labels of all internal faces (the faces
not belonging to the boundary).  The three products run over the the
2-simplices $f$, the 3-simplices $e$ and the 4-simplices $v$ of the
triangulation, respectively (the choice of the letter refers to
corresponding dual elements: faces, edges and vertices.)  The labels
$(\rho^e_{1},\ldots,\rho^e_{4})$ are the ones of the four 2-simplices
adjacent to the 3-simplex $e$.  The labels
$(\rho^v_{1},\ldots,\rho^v_{10})$ are the ones of the ten 2-simplices
adjacent to the 4-simplex $v$.

\begin{figure}
\centerline{{\psfig{figure=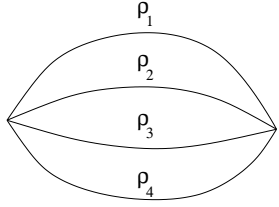,height=3cm}}}
\bigskip \caption{The $\Theta_{4}$ spin net.}
\label{Figure:theta}
\end{figure} 

\begin{figure}
\centerline{{\psfig{figure=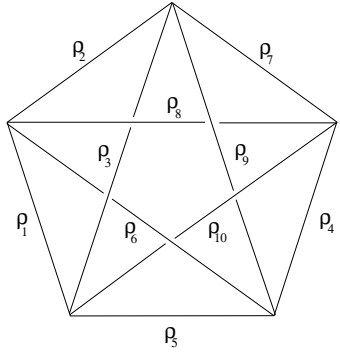,height=5cm}}}
\bigskip \caption{The $I_{10}$ spin net.}
\label{Figure:I10} 
\end{figure}

The functions $\Theta_{4}$ and $I_{10}$ are defined as traces of
recombination diagrams for the balanced representations, regularized
as explained in \cite{Barrett:2000qw}.  The function $\Theta_{4}$ is
given by the diagram in Figure 1.  It was discovered to play a role in
the model in \cite{Perez:2001ec} and its evaluation is in
\cite{Barrett:2000qw}.  The function $I_{10}$ is given by the diagram
in Figure 2.  These traces are called relativistic spin networks.  As
shown in \cite{Barrett:2000qw}, these relativistic spin networks can
be explicitly expressed as multiple integrals on the upper sheet $H$
of the $2$-sheeted hyperboloid in Minkowski space.  To this purpose,
we define the projector kernel
\begin{equation}
K_{\rho}(x,y)=\frac{\sin(\rho\ d(x,y))}{\rho\ \sinh(d(x,y))}
\label{kernel}
\end{equation} 
where $d(x,y)$ is the hyperbolic distance between $x$ and $y$.  Then
the trace of a recombination diagram is given by a multiple integral
of products of $K$'s.  More precisely, by one integral over $H$ per
each node, of the product on one kernel per each link.  The integral
is then normalized by dropping one of the integrations.  By Lorentz
symmetry, the result is independent of the point not integrated
over.  Thus in particular $\Theta_{4}$ and $I_{10}$ are given by
\begin{equation}
    \label{theta}
\Theta_4(\rho_{1},\ldots,\rho_{4})= 
\frac1 {2\pi^2} 
\int_{{H}}K_{\rho_1}(x,y) \ldots K_{\rho_{4}}(x,y)\,\dd y 
\end{equation}
and
\begin{equation}
I_{10}(\rho_{1},\ldots,\rho_{10})= 
\frac1 {2\pi^2} 
\int_{{H}^4}\prod_{i\le j=1,5}
K_{\rho_{ij}}(x_{i},x_{j})\ 
\dd x_{1}\dd x_{2}\dd x_{3}\dd x_{4}.  
\label{I10}
\end{equation}
Equations (\ref{one}--\ref{I10}) define the
state sum completely.  For a four dimensional manifold with boundary,
(\ref{one}) gives a function of the boundary labels.  These functions
can be interpreted as three-geometry to three-geometry transition
amplitudes, computed to a certain order in a perturbative expansion. 
They can be viewed as the (in principle) observable quantities of a
quantum theory of gravity, as explained in \cite{Perez:2001gj}.

Each term in the sum (\ref{one}) is a multiple integral on an
unbounded domain.  The functions $\Theta_{4}$ and $I_{10}$ were shown
to be bounded in \cite{Barrett:2000qw} and \cite{Baez:2001fh}. 
Therefore convergence is a question of sufficiently rapid decay at
infinity.  Since we are assuming a nondegenerate triangulation, each
2-simplex in (\ref{one}) appears in at least three distinct
3-simplices and at least three distinct 4-simplices.  Therefore, by
power counting, we need a combined power law decay at infinity adding
to more than 1 in the combination of the $\Theta_{4}$ and $I_{10}$
factors for each $\rho$ separately.  As we show below, the
$\Theta_{4}$ factor has a power law decay with exponent 3/4, while the
$I_{10}$ has a decay with exponent 3/10.  The first exponent seems to
be sharp, the second can probably be strengthened but not by very
much.  We now supply a rigorous proof of finiteness.

\section{The proof}

We begin with two results by Baez and Barrett in \cite{Baez:2001fh},
which we state without proof.
\begin{lemma}
    \label{bound}
    (Baez-Barrett) $\Theta_{4}$ and $I_{10}$ are bounded.
\end{lemma} 
This follows immediately from Theorems 1, 2 and 3 of
\cite{Baez:2001fh} \begin{lemma} (Baez-Barrett)
\label{inductive} 
If $n \ge 3$, the integral 
\[    
J(x_{1},\ldots,x_{n}) = \int_{H} \dd x \, |K_{\rho_1}(x,x_1)
K_{\rho_2}(x,x_2) \cdots K_{\rho_n}(x,x_n)|
\] 
converges, and for any $0<\epsilon < 1/3$ there exists a constant 
$C > 0$ such that for any choice of the points $x_1,\dots,x_n$,  
\[ 
J \le C\,  
\exp\left(-\frac{n-2-n\epsilon}{n(n-1)}\sum_{i<j}r_{ij}\right) 
\] 
where $r_{ij}$ is the hyperbolic distance $d(x_i,x_j)$  between $x_i$ and
$x_j$.
\end{lemma} 
This result, (Lemma 5 in reference \cite{Baez:2001fh}) is one of the
fundamental tools in the proof of Lemma {\ref{bound}}.

\begin{lemma} The $\Theta_4$ relativistic spin network satisfies the
following bound
\begin{equation}
        |\Theta_4| \le \frac {1}{\rho_1\rho_2\rho_3}
    \label{eq:lemma2}
\end{equation}
for any arbitrary triple $\rho_1\rho_2\rho_3$.
\end{lemma}

\begin{proof}
From the fact that $|K_{\rho}|\le 1$ we have that $|\Theta_4|\le
|\Theta_3|$, where $\Theta_3$ corresponds to the evaluation of the
spin network obtained by dropping an arbitrary link from the original
one.  
\begin{equation}
    \Theta_3= \frac1 {2\pi^2} \int_{{H}} K_{\rho_1}(x,y)
    K_{\rho_2}(x,y) K_{\rho}(x,y)\,\dd y
\label{tete}
\end{equation}
As it is shown in \cite{Barrett:2000qw}, 
\begin{eqnarray}
\nonumber \Theta_3&=& \frac{2}{ \pi \rho_1\rho_2\rho_3}\int_0^\infty
\frac{\sin \rho_1 r\sin \rho_2 r\sin \rho_3 r}{\sinh r}\,\dd r \\
\nonumber &=&\frac1{4 \rho_1\rho_2\rho_3} \bigl(\tanh(\frac\pi2(
\rho_1+\rho_2-\rho_3))+ \tanh(\frac\pi2(\rho_3+\rho_1-\rho_2))\\
&&\nonumber + \tanh(\frac\pi2(\rho_2+\rho_3-\rho_1))-
\tanh(\frac\pi2(\rho_1+\rho_2+\rho_3))) \bigr)
\end{eqnarray}
From this we have 
\begin{equation}
    |\Theta_{3}| \le \frac 1 { \rho_1\rho_2\rho_3}
\end{equation}
and (\ref{eq:lemma2}) follows. 
\end{proof}
\begin{corollary}\label{four}
For any subset of $\kappa$ elements $\rho_1 \dots \rho_{\kappa}$ out
of the corresponding four representations appearing in $\Theta_{4}$
the following bounds hold \begin{equation}\nonumber \left| \Theta_4
\right| \le \frac {C_{\kappa}} {\left( \prod \limits^{\kappa}_{i=1}
\rho_i \right)^{\alpha_{\kappa}}} \ \ \ {where} \ \ \
\alpha_{\kappa}=\left\{ \begin{array}{ccc} 1\ \ { for}\ \ \kappa \le 3
\\
 \frac{3}{4}\ \ { for}\ \  {k=4}
\end{array}\right.. 
\end{equation}
for some positive constant $C_{\kappa}$.
\end{corollary}

\begin{proof}
The case $\kappa=3$ corresponds to equation (\ref{tete}).  For $\kappa
< 3$ we observe that in the definition of $\Theta_3$ in (\ref{tete})
we can obtain a bound containing $\kappa$ different $\rho$'s in the
denominator by bounding $3-\kappa$ of the three $K_{\rho}$'s by $\frac
r {\sinh r}$.  For the case $\kappa=4$ we can write four inequalities
as in the previous Lemma choosing different triplets.  Multiplying the
four inequalities each representation appears repeated three times so
we obtain the exponent $\frac 3 4$ in the bound.
\end{proof}

\begin{lemma}\label{6j}
The tetrahedron amplitude 
\begin{eqnarray}   \label{Ghandi}
    I_{6} &=& \int_{{H}^3} \dd x_2 \,\dd x_3 \,\dd x_4 \; 
K_{\rho_{12}}(x_1,x_2) K_{\rho_{13}}(x_1,x_3) K_{\rho_{14}}(x_1,x_4) \\  
&& \nonumber \qquad \qquad  
K_{\rho_{23}}(x_2,x_3) K_{\rho_{24}}(x_2,x_4) K_{\rho_{34}}(x_3,x_4). 
\end{eqnarray} 
(see Figure 3) satisfies 
\begin{equation}
    I_6 \le \frac{K}{\rho_1\rho_2\rho_3},
    \label{eq:lemma3}
\end{equation}
for a constant $K$, and any choice of three non vanishing colorings
$\rho_1,\rho_2$, and $\rho_3$ in the same triangle.
\end{lemma}

\begin{figure}
\centerline{{\psfig{figure=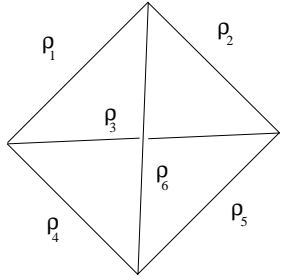,height=4cm}}}
\bigskip \caption{The $I_{6}$ spin net.}
\label{I6}
\end{figure} 

\begin{proof}
We study the integral 
\begin{eqnarray}   
    I &=& \int_{{H}^3} \dd x_2 \,\dd x_3 \,\dd x_4 \; 
|K_{\rho_{12}}(x_1,x_2) K_{\rho_{13}}(x_1,x_3) K_{\rho_{14}}(x_1,x_4) \\  
&& \nonumber \qquad \qquad  
K_{\rho_{23}}(x_2,x_3) K_{\rho_{24}}(x_2,x_4) K_{\rho_{34}}(x_3,x_4)|. 
\end{eqnarray} 
for any choice of numbers
$\rho_{ij} \ge 0$ for $1 \le i < j \le 4$ and a point $x_1 \in {H}$. 
First we integrate out $x_4$ using Lemma \ref{inductive}, obtaining 
\begin{eqnarray}\label{interm}\nonumber &&
I \le C \int_{{H}^2} \dd x_2 \,\dd x_3 \;  
e^{-\frac 1 6 (1 - 3\epsilon)(r_{12} + r_{13} + r_{23})} \\
&&\ \ \ \ \ \ \  \ \ \ \ \ \ \ \ \ \ \ \ \ \
|K_{\rho_{12}}(x_1,x_2) K_{\rho_{13}}(x_1,x_3) K_{\rho_{23}}(x_2,x_3)|
\end{eqnarray} 
where $r_{ij} = d(x_i,x_j)$. We can bound the previous expression
by
\begin{equation}
\label{inti} 
I \le \frac{C}{\rho_{12}\rho_{13}\rho_{23}} \int_{{H}} \dd
x_2 \frac{e^{-\frac 1 6 (1 - 3\epsilon)r_{12}}}{\sinh r_{12}}\,
\int_{{H}}\dd x_3 \; \frac{e^{-\frac 1 6 (1 - 3\epsilon)(r_{13} +
r_{23})}}{\sinh r_{13} \sinh r_{23}}. 
\end{equation}
Lets concentrate on the $x_3$ integration. In order to do so
we use a coordinate system in which two of the coordinates are  
\[ k  = \frac 1 2 (r_{13} + r_{23}), \qquad  
   \ell = \frac 1 2 (r_{13} - r_{23}), \] 
while the third is the angle $\phi$ between $x_3$ and a given plane 
containing the geodesic between $x_1$ and $x_2$.  The ranges of these 
coordinates are 
\[   r/2 \le k < \infty, \qquad   
     -r/2 \le \ell \le r/2 , \qquad     
     0 \le \phi < 2 \pi , \] 
where we set $r = r_{12}$.  
In terms of this coordinates the measure $\dd x_3$ becomes
\[ \dd x_3 =  
2\,\frac{\sinh r_{13} \, \sinh r_{23}} {\sinh r} \; \dd k \, \dd \ell \,  
\dd \phi. \]    
In terms of this coordinates (\ref{inti}) becomes 
\begin{eqnarray} 
\nonumber && I \le \frac{2 C}{\rho_{12}\rho_{13}\rho_{23}} \int_{{H}} \dd x_2 
\frac{e^{-\frac 1 6 (1 - 3\epsilon)r }}{\sinh r }\, 
\int_0^{2 \pi} \dd \phi \int_{r/2}^\infty  
\dd k \int_{-r/2}^{r/2} \dd \ell \;  
\frac{e^{-\frac 1 3 (1 - 3\epsilon)k}}{\sinh r}. 
\end{eqnarray}
Finally if we put in the form of the measure $\dd x_2$, i.e. $\dd
x_2=\sinh^2 r \dd r \dd \Omega$ (where $\dd \Omega$ is the measure of
the unit sphere), we can complete the integration to obtain the sought
for bound, namely
\begin{eqnarray}
\nonumber  I &\le& \frac{16 \pi^2 C}{\rho_{12}\rho_{13}\rho_{23}}
\int_0^{\infty} \dd r {e^{-\frac 1 6 (1 - 3\epsilon)r }}\,
\int_{r/2}^\infty \dd k \; r {e^{-\frac 1 3 (1 - 3\epsilon)k}}\\
& =& \frac{16 \pi^2 C}{\frac 1 3 (1 -
3\epsilon){\rho_{12}\rho_{13}\rho_{23}}} \int_0^{\infty} \dd r\ r\ 
{e^{-\frac 1 3 (1 - 3\epsilon)r }},
\end{eqnarray}
which concludes the proof.
\end{proof}
\begin{lemma}\label{yo}
The 4-simplex amplitude $I_{10}$ satisfies the following bound:
\[I_{10}\le \frac{K}{\rho_1 \rho_2 \rho_3},\]
for some constant $K$.
\end{lemma}
\begin{proof}
The $4$-simplex amplitude $I_{10}$ corresponds to introducing four
additional $K_{\rho}$ in the multiple integral ({\ref{Ghandi}})
together with an additional integration corresponding to the four new
edges and the additional vertex respectively.  Using Lemma
{\ref{inductive}} this additional integration can be bounded by a
constant, so that after using Lemma \ref{6j} we have \[I_{10}\le
\frac{K}{\rho_1 \rho_2 \rho_3}\] for any arbitrary triple $\rho_1
\rho_2 \rho_3$ in the same triangle.
\end{proof}

\begin{lemma}\label{yo1}
$I_{10}$ satisfies also the following bounds: \[I_{10}\le
\frac{K_1}{\rho_1\rho_{2}}, \ \ \ I_{10}\le \frac{K_2}{\rho_1}, \]
where $K_1$, and $K_2$ are constant and $\rho_1$ and $\rho_2$
are in the same triangle.
\end{lemma}
\begin{proof}
We observe that a different bound can be obtained for $I_6$ containing
respectively two or one representations in the denominator if we bound
either two or one of the three $K_{\rho}$ in (\ref{interm}) by $\frac
r {\sinh r}$ instead of just taking absolute value.  The integration
on the right still converges (see \cite{Baez:2001fh}).
\end{proof}
Using Lemma \ref{yo} and \ref{yo1} it is easy to prove the following
corollary.

\begin{corollary}\label{ten} For any subset of $\kappa$ elements
$\rho_1 \dots \rho_{\kappa}$ out of the corresponding ten
representations appearing in $I_{10}$ the following bounds hold
\begin{equation}\nonumber \left| I_{10} \right| \le \frac {K_{\kappa}}
{\left( \prod \limits^{\kappa}_{i=1} \rho_i \right)^{\frac 3 {10}}}.
\end{equation}
for some positive constant $K_{\kappa}$.
\end{corollary}

\begin{theorem}
Given a non singular triangulation, the state sum partition function
${\cal Z}$ is well defined, i.e., the multiple integral in (\ref{one})
converges.
\end{theorem}
\begin{proof}
We divide each integration region ${\mathbb R}^+$ into the intervals $[0,1)$,
and $[1,\infty)$ so that the multiple integral decomposes in a finite
sum of integrations of the following types:
\begin{enumerate}
\item[i.] All the integrations are in the range $[0,1)$.  We denote this
term $T(F,0)$, where $F$ is the number of 2-simplices in the
triangulation.  This term in the sum is finite by Theorem
{\ref{bound}}.  
\item[ii.] All the integrations are in the range
$[1,\infty)$.  This term $T(0,F)$ is also finite since using Corollary
\ref{four} and {\ref{ten}} for $\kappa=4$, and $\kappa=10$
respectively we have 
\[ 
T(0,F) \le \prod_{f} \int
\limits^{\infty}_{\rho_{f}=1} \dd \rho_f \ \ \rho_{f}^{2-\frac3 4
{n_e} - \frac3 {10} {n_v}} \le \left(\ \int
\limits^{\infty}_{\rho_{f}=1} \dd \rho_f \ \ \rho_{f}^{-\frac{42}{40}}
\right)^F < \infty
\] 
\item[iii.]  $m$ integrations in $[0,1)$, and $F-m$ in $[1,\infty)$. 
In this case $T(m,F-m)$ can be bounded using Corollaries (\ref{four})
and (\ref{ten}) as before.  The idea is to choose the appropriate
subset of representations in the bounds (and the corresponding values
of $\kappa$) so that only the $m-F$ representations integrated over
$[1,\infty)$ appear in the corresponding denominators.  Since this is
clearly possible, the $T(m,F)$ terms are all finite.
\end{enumerate}
We have bounded ${\cal Z}$ by a finite sum of finite terms which
concludes the proof.
\end{proof}

\section{Conclusion}

Given the history of attempts to quantize general relativity, the
finiteness of the Lorentzian state sum is remarkable, and came to us
as a surprise.  The model is the result of a number of choices made
from physical or geometrical arguments.  Altering one of those choices
seems to generally have the effect of destroying the finiteness.  An
example is the uncertainty, in the older 15J formulation, as to
whether to sum independently on the internal spins in tetrahedra or to
require them to be equal.  The version of the model studied here has
$\Theta_{4}$ terms which can be thought of as sums of 6J terms, or
alternatively as diagrams which have the effect of forcing the two
internal labels to be equal.  Thus finiteness seems to impose a choice
here, the other version would almost certainly diverge.  Similarly,
without the balanced constraint of \cite{Barrett:1998gw}, the model
would also, most likely, be divergent.  That constraint can be deduced
from geometrical thinking or from the Plebansky formulation of general
relativity; both derivations are unrelated to the representation
theory which ensued, so it is interesting that it plays a critical
role in finiteness.

In order to turn the model into a complete theory, it is necessary to
handle the limit as the number of simplices in a triangulation goes to
infinity.  It is possible to imagine several plausible approaches to
this.  The simplest would be to cut off the number of triangulations
and regularize, hoping for a good limit.  The conjecture of quantum
self censorship of \cite{Crane:2000fj} might play a role, either in
this model or in an extended one with matter terms added, in ensuring
that such a limit would exist; the intuition being that any new
information in a sufficiently large triangulation would fall into its
Schwarzschild radius, and hence not affect physically observable
quantities. 

An alternative to just studying the limit of larger triangulations is
to sum over triangulations, or suitably extended triangulations
\cite{DePietri:2000bx,Reisenberger:2000fy,Reisenberger:2001zc}.  From
this perspective, the state sum we have studied here is a term in the
Feynman expansion of an auxiliary field theory \cite{Perez:2001ec}. 
The field theory fixes the relative weights in the sum over extended
triangulations, which includes also a sum over spacetime topologies. 
In this perspective, the result in this paper is the finiteness of
Feynman integrals.  In such a context, however, singular
triangulations appear as well, an issue not addressed here. 
Finiteness on singular triangulations requires us to get a power decay
law above 3/4 for the $I_{10}$ integrals.  We do not know if this is
possible or not.  The question of including manifolds with conical
singularities, or equivalently, more general 2-complexes, is related
to this.  The possibility of discrete geons is therefore open.

Another question which we think deserves to be investigated is the
finiteness of the variant of the model considered in
\cite{Perez:2000ep}, in which timelike as well as spacelike balanced
representations are used, and in which the discreteness characteristic
of the canonical theory \cite{Rovelli:1997} reappear.

More generally, the results presented here emerge from the comparison
of two different ways of viewing spinfoam models: the quantum
geometric and field theoretic ones.  We believe that comparing this
approach with the techniques and the results from the other approaches
to the construction of spinfoam models, such as the ones of
Reisenberger \cite{Reisenberger:1997sk}, Freidel and Krasnov
\cite{Freidel:1999pt} and Iwasaki \cite{Iwasaki:2000xr}, is likely to
be productive as well.

The suggestion that quantum geometry is in some sense discrete is an
old idea in quantum gravity: it can be traced all the way to Einstein
\cite{Stachel}.  We view the categorical algebraic elegance of the
current model in which such discreteness is realized, and its close
relationship to TQFTs, as particularly attractive.  The structural
similarity between categorical state sums and the Feynman vacuum,
emphasized in \cite{Crane:2000fj}, is also suggestive.  Because of the
finiteness proof given here, it is now possible to explore the
consequences of these ideas and the physical content of this model by
exact calculation.

\centerline{-----------------}

This work was completed with support from NSF Grants PHY-9900791 and 
DMS-9504423 and the Andrew Mellon Predoctoral Fellowship.

\end{document}